\pdfoutput=1
      [1999/12/01 v1.4c Il Nuovo Cimento]
\documentclass{cimento}

\usepackage{graphicx}  
\usepackage{epsfig}
\title{To be or not to be: Higgs impostors at the LHC}
\author{A.~De R\'ujula\from{ins:evil}}
\instlist{
  \inst{ins:evil} Instituto de F\'isica Te\'orica, Universidad Aut\'onoma \& CIEMAT (Madrid, Spain),\\
  CERN (Switzerland), Boston University (USA).}
\PACSes{\PACSit{14.80.Bn}{14.80.Cp}}

\begin{document}

\maketitle

\begin{abstract}
Consider the day when an invariant mass peak, roughly compatible with ``the Higgs",
begins to emerge, say at the LHC, ...~and may you see that day. 
There will be a difference between discovery and 
scrutiny. The latter would involve an effort to ascertain what it is, or is not,
that has been found. 
It turns out that the two concepts are linked: 
Scrutiny will naturally
result in deeper knowledge -- is *this* what you were all looking for? --
but may also speed up discovery.

\end{abstract}

\section{Introduction}

\begin{figure}[htb!]
\begin{center}
\includegraphics[width=1.0\textwidth]{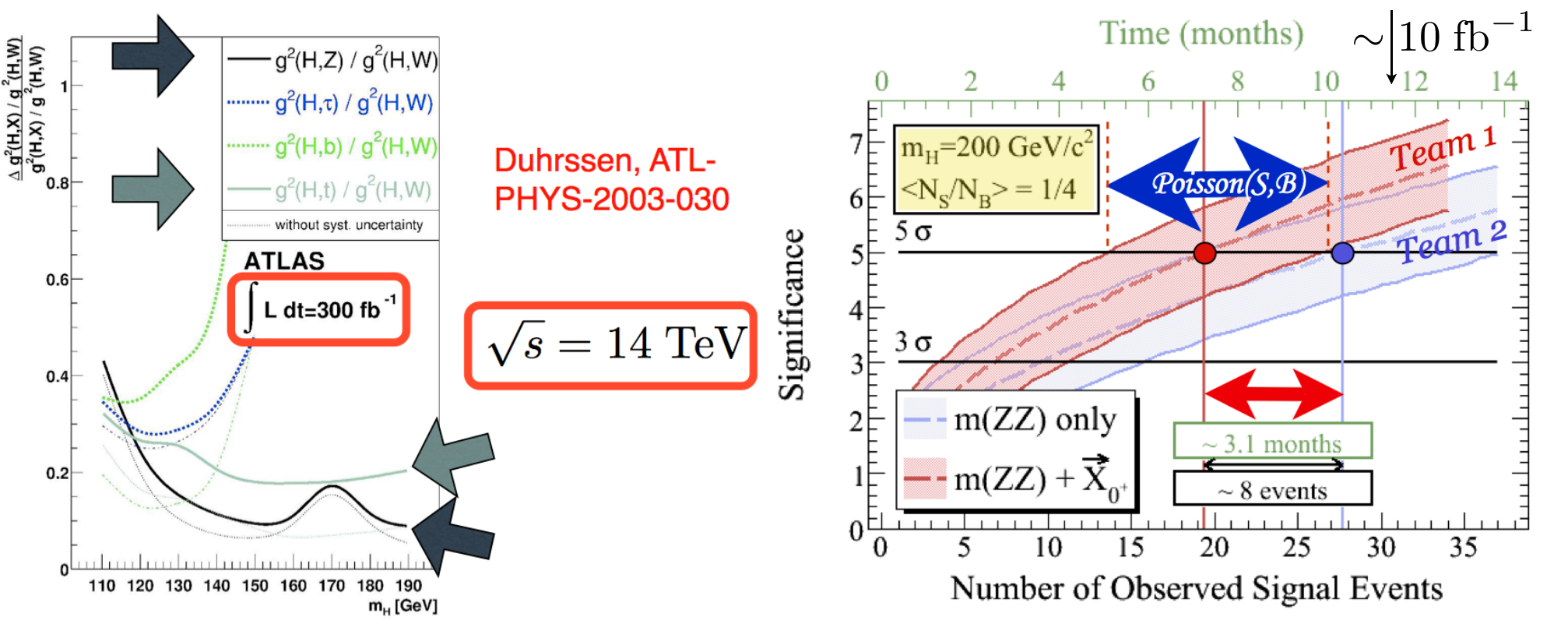}
\caption{Left: Fractional precision on the measurable ratios of branching ratios 
for SM $H$ decays into $W$, $Z$, $t$ 
and $\tau$ pairs as functions of $M_H$. 
Right: An example of discovery and scrutiny plot
of a SM scalar with $M_H=200$ GeV, not specially chosen for effect.
$\vec X_0$ as in Fig.~\ref{CMVariables}.
\label{BRsDisco}}
\end{center}
\end{figure}

Let the single missing scalar of the Standard Model (SM) be called ``the Higgs", to stick to 
a debatable misdeed. Because the idea is so venerable, one may have grown insensitive to
how special a Higgs boson would be.  Its quantum numbers must be those
of the vacuum, which its field permeates. The boson itself would be the vibrational quantum
*of* the vacuum, not a mere quantum *in* the vacuum, or in some other substance.
The couplings of the Higgs to quarks and leptons are proportional to their masses. So are
its couplings to $W^\pm$ and $Z$, a fact that, within the SM,
is in a sense verified.  A significantly precise direct measurement of the Higgs couplings to 
fermions is not an easy task. Even for the heaviest of them, the top quark, the required integrated
luminosity is large, as illustrated by the ATLAS collaboration on the left of Fig.~\ref{BRsDisco}.

In the past, given a newly discovered particle, one had to figure out its 
$J^{PC}$ quantum numbers 
(or its disrespect of the super-indexed ones) to have it appear
in the Particle Data Book. Publication in the New York
Times was not considered that urgent, nor was it immediate for bad news. 
Times have changed. Yet, 
two groups~\cite{ref:Others,ref:HLL} have thoroughly studied
the determination of the quantum numbers and coupling characteristics
of a putative signal at the LHC, that could be the elementary scalar of the SM,
 or an {\it impostor} thereof, both dubbed $H$ here. 
The ``golden channel" for this exercise is $H \to (ZZ$ or $ZZ^{*})\to
\ell^+_1\ell^-_1\ell^+_2\ell^-_2$, where $\ell^\pm_{1,2}$ is an
$e$ or a $\mu$, and $Z^{*}$ denotes that, for $M_H<2\,M_Z$, one of the $Z$s is
 ``off-shell". For a review of previous work on the subject, see
e.g.~\cite{ref:Djouadi}.

To be realistic (?) let me consider two competing teams. They are 
working at a $pp$ collider of energy $\sqrt{s}=10$ TeV, luminosity $10^{33}$
cm$^{-2}$ s$^{-1}$ and Snowmass factor of 3 (on average, things work well
1/3 of the time). The SM is correct, $M_H=200$ GeV and the
estimates of signals and backgrounds are reliable. As the number of events increases,
Team 2 would then gather evidence for an $M_{ZZ}$ peak
at the rate shown on the right of Fig.~\ref{BRsDisco}. Team 1 is additionally
checking that, indeed, the object has $J^{PC}=0^{++}$. T1 reaches
``discovery" ($5\sigma$ significance) some three months before T2.
The horizontal error bars, dominated by  fluctuations in the
expected background, tell us that the two teams are *only*
$1\sigma$ apart (iff from two different experiments!). But that means the
probability of T1 (from experiment $A$) being 3 months ahead of T2 
(from experiment $B\neq A$) is $\sim 66$\% ($\sim 100$\% for $B=A$). The odds for winning
with dice, if your competitor lets you win for 4 out of the 6 faces are also 66\%.
If the stakes are this high, would you not play? 
It is interesting to compare the two $H$-identity-revealing 
integrated luminosities in Figs.~\ref{BRsDisco}, more so
since event numbers on its right
refer to the chain $H\to ZZ \to e^+e^-\,\mu^+\mu^-$ and are
approximately quadrupled when all $4\ell$ channels are considered.

Standard signal and background cross sections times branching ratios, $\sigma\times B$,
were used in Fig.~\ref{BRsDisco}. In discussing $H$ impostors we accept
that they should not be distinguished from a SM $H$ on $\sigma\times B$ grounds, which, for
all impostors, are hugely model-dependent.

\section{Methodology}
The technique to be used to measure $J^{PC}$ for a putative $H$ signal has some pedigree.
Its quantum-mechanical version (called nowadays the ``matrix element" method) capitalizes 
on the entanglement of the two $Z$ polarizations and dates back at least to the first (correct)
measurements of the correlated $\gamma$ polarizations in parapositronium ($0^{-+}$)
decay~\cite{ref:Wu}. The technique is even older, as it actually consists in comparing theory and observations. The art is in exploiting {\it a maximum of the information} from both sides.

\begin{figure}[htb!]
\begin{center}
\includegraphics[width=0.6\textwidth]{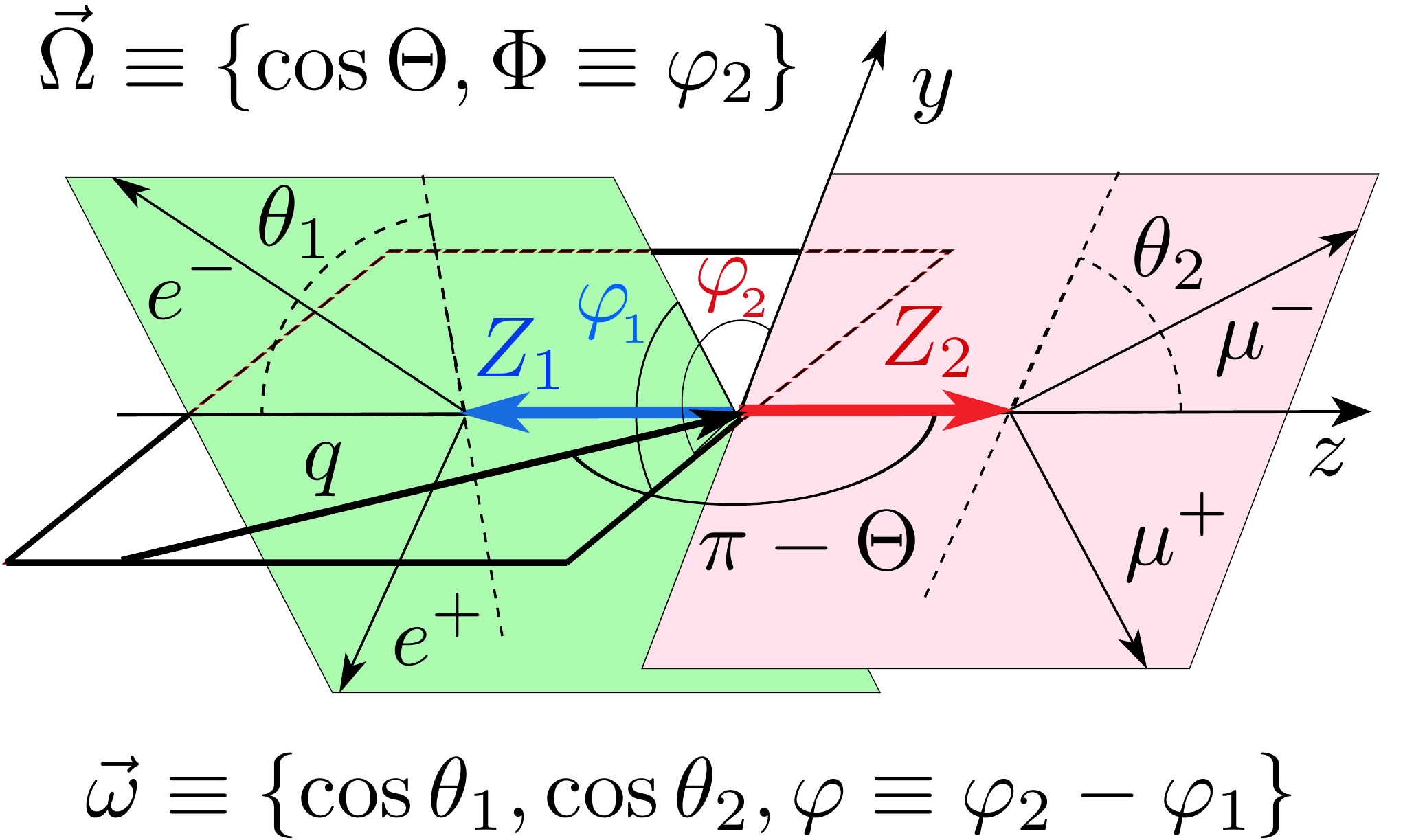}
\caption{The angles of $ZZ$ pair-production and leptonic decay. 
$\vec X_0 \equiv \{ \vec \Omega, \vec \omega\}$.
\label{CMVariables}}
\end{center}
\end{figure}

\begin{figure}[tbp]
\begin{center}
\includegraphics[width=0.3\textwidth]{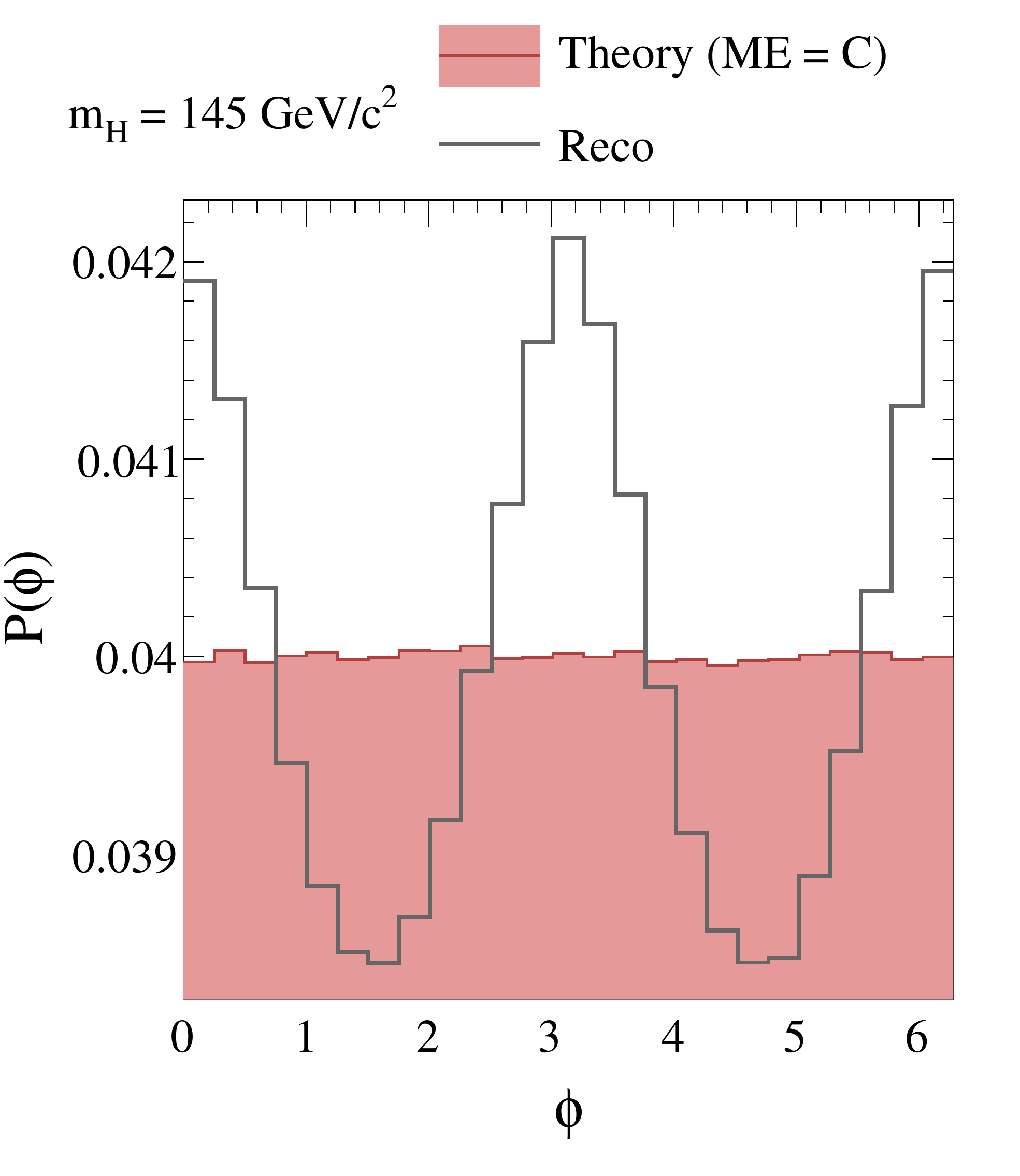}
\includegraphics[width=0.3\textwidth]{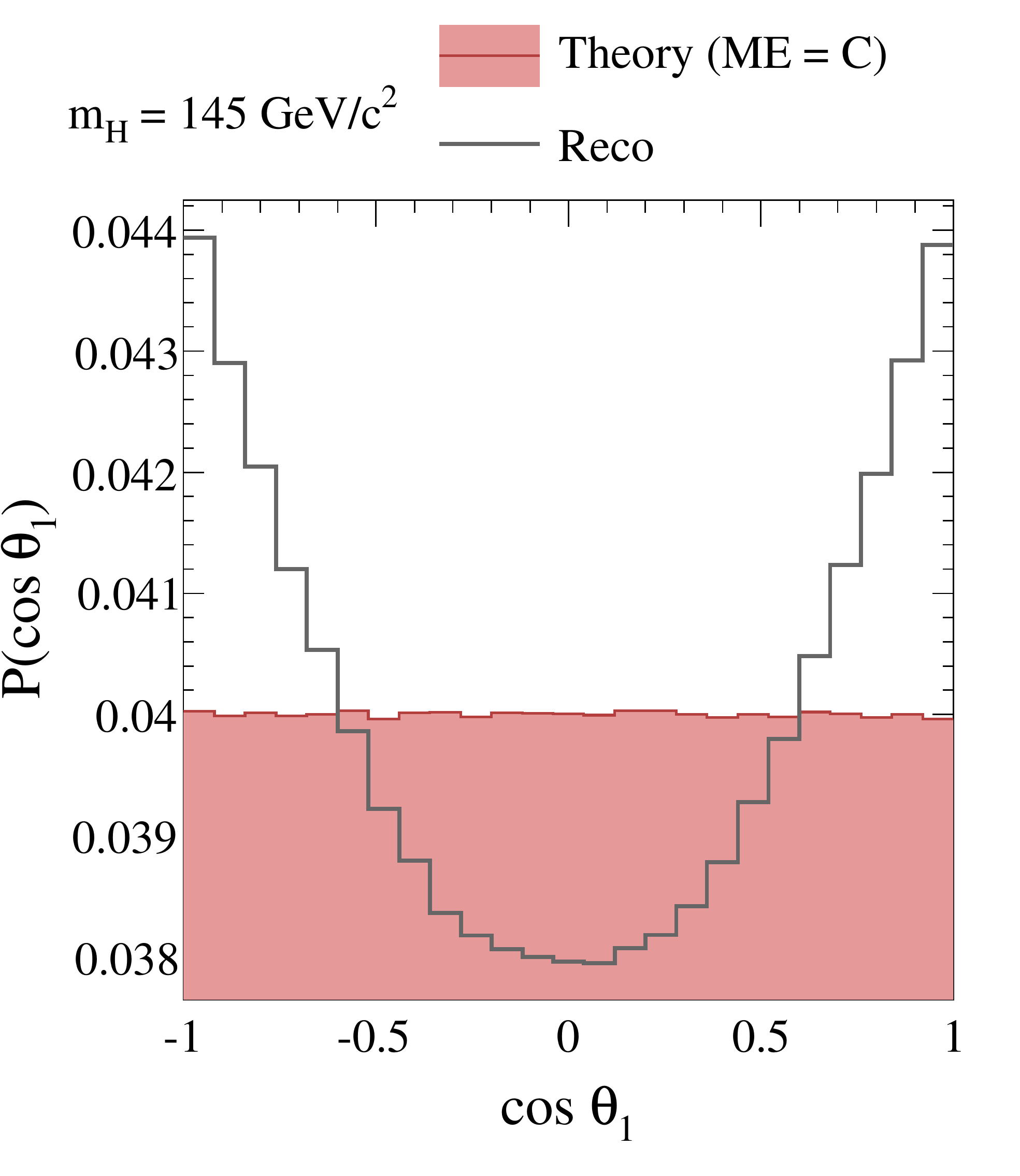}
\includegraphics[width=0.3\textwidth]{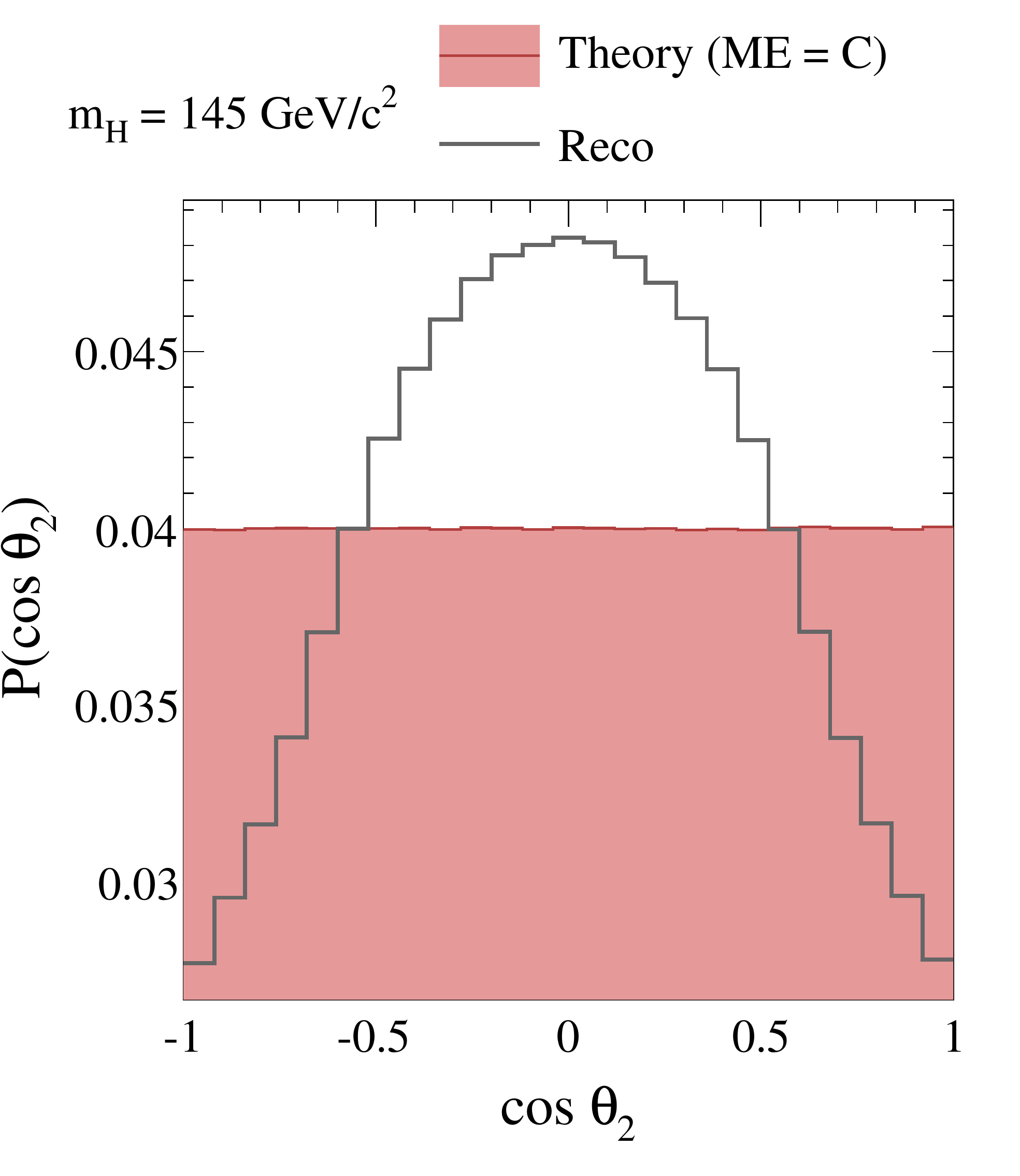}\\
\includegraphics[width=0.3\textwidth]{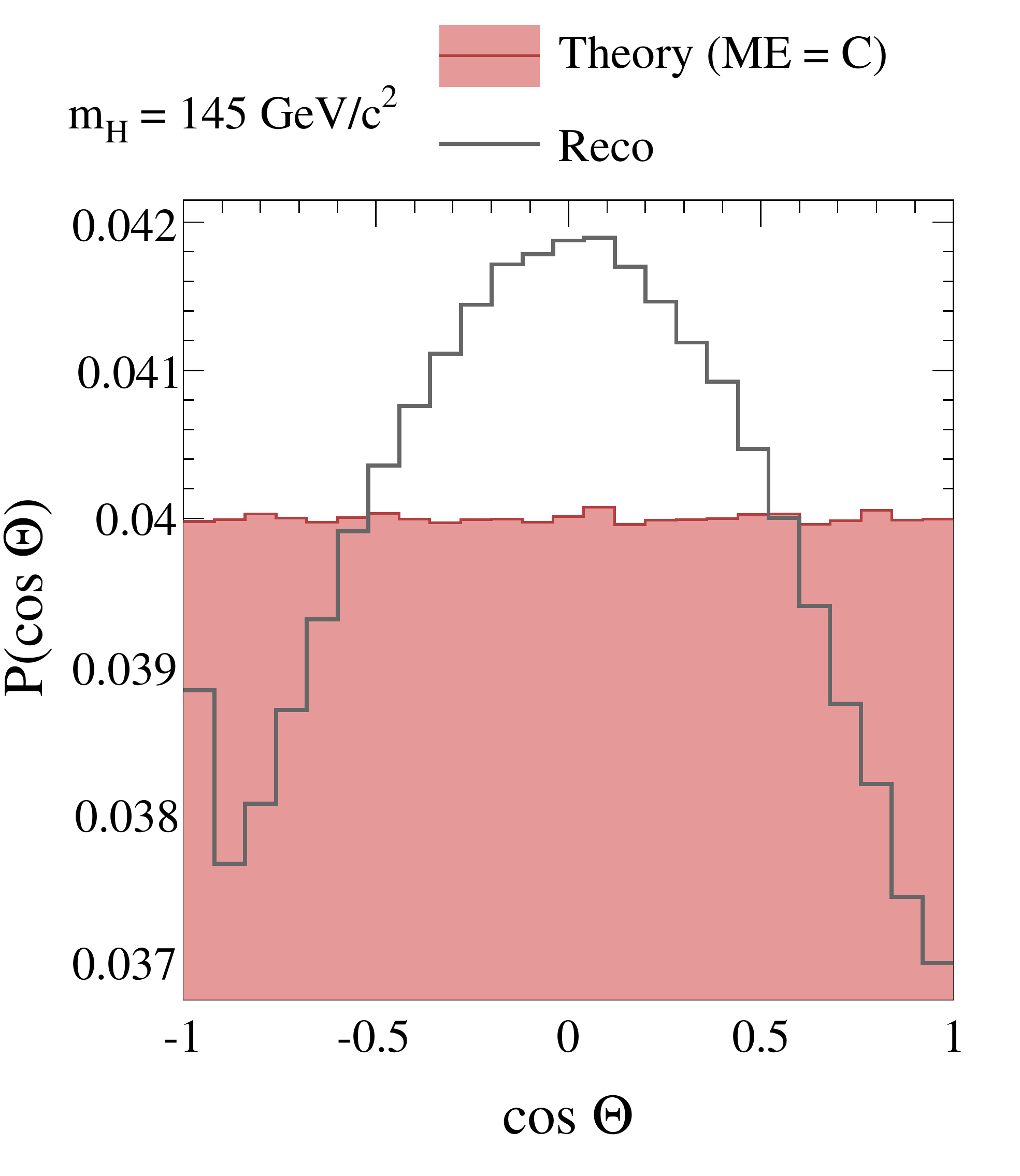}
\includegraphics[width=0.3\textwidth]{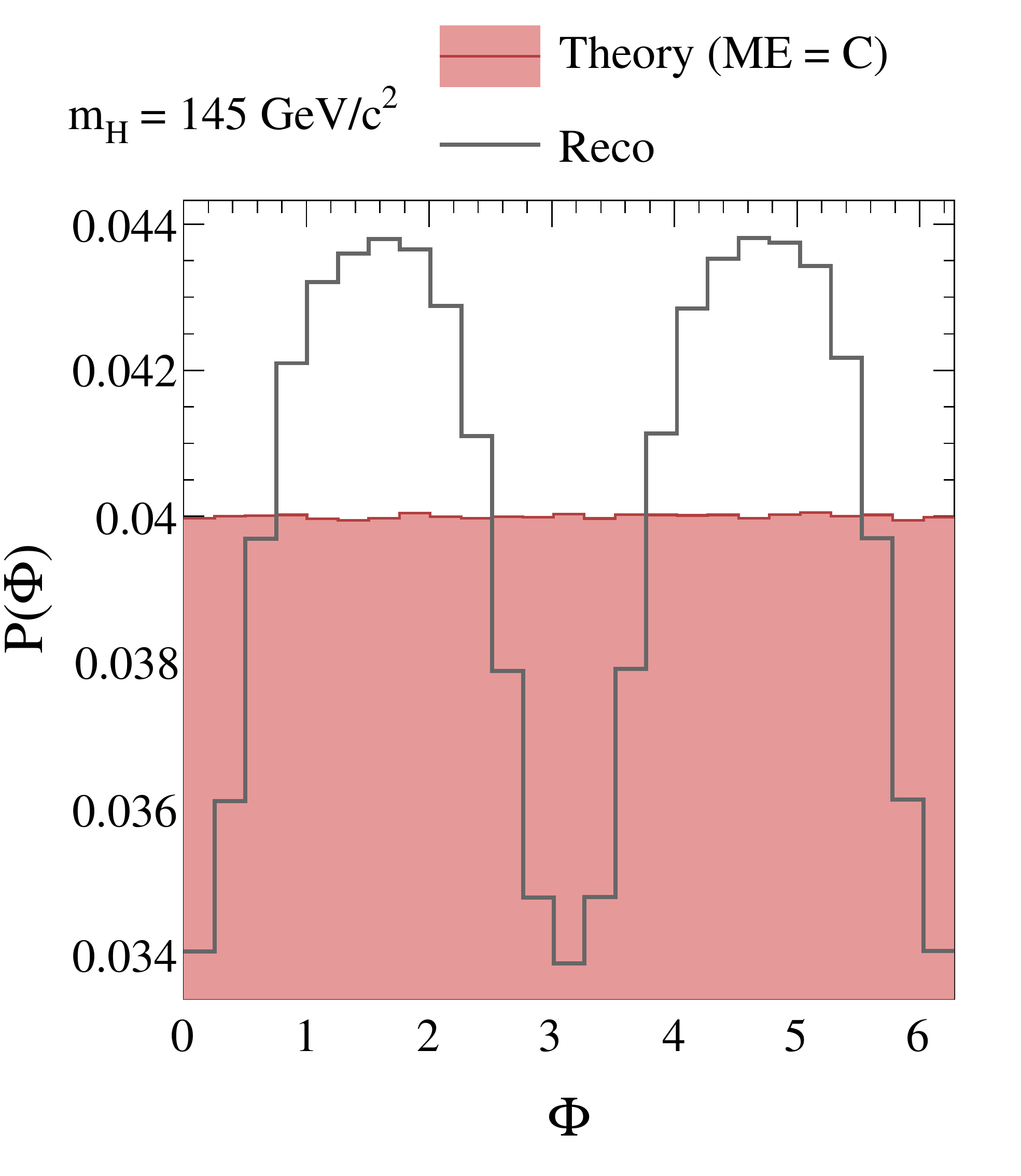}
\includegraphics[width=0.3\textwidth]{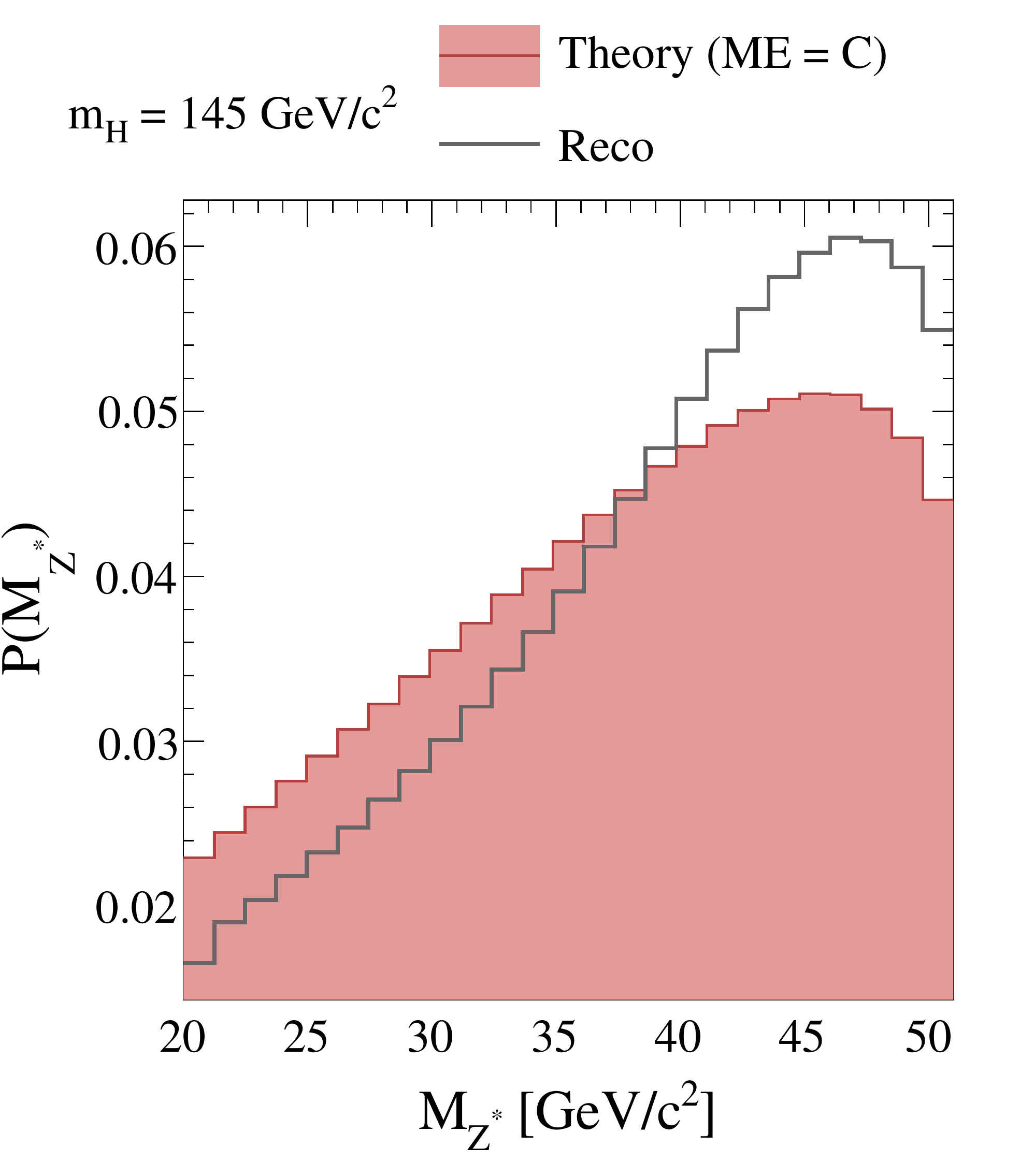}
\caption{Detector-shaping effects at
$M_H=145$ GeV, for all relevant angles and $M^*$. The trigger and energy thresholds, resolutions and angular coverage are those of a ``typical" detector.
\label{Shaping}}
\end{center}
\end{figure}
%


The event-by-event information on the channel at hand is very large, some of it is illustrated 
in Fig.~\ref{CMVariables}, for the decay chain $H\to ZZ \to e^+e^-\,\mu^+\mu^-$,
with $H$ brought to rest. The angular variables $\vec\Omega$ describe $Z$-pair production
relative to the annihilating $gg$ or $q\bar q$ pair.
The variables $\vec\omega$ are the $Z$-pair decay 
angles. For fixed $\vec\Omega$, $\vec\omega$, and $M^*$ 
(the mass of a lepton pair if its parent $Z$ is off-shell) that is all there is: none less than six 
beautifully entangled variables ($M[4\ell]$ is also measured event by event,
$M_H$ is traditionally extracted from a fit to the $M[4\ell]$ distribution).

Real detectors have limited coverage in angles and momenta, they ``mis-shape"
the theoretical distributions in the quantities just described. An example for a
realistic detector and an unrealistic flat expectation is illustrated on the right of
Fig.~\ref{Shaping}. For an $H$ with $J=0$, the distribution 
in $\vec\Omega$ is flat, so that its inclusion (in this case) would seem
like an overkill. Not so! detector-shaping effects and the correlations between the
angular variables conspire to make the use of the full machinery a necessity~\cite{ref:HLL}.

\begin{figure}[htb!]
\begin{center}
\includegraphics[width=1.0\textwidth]{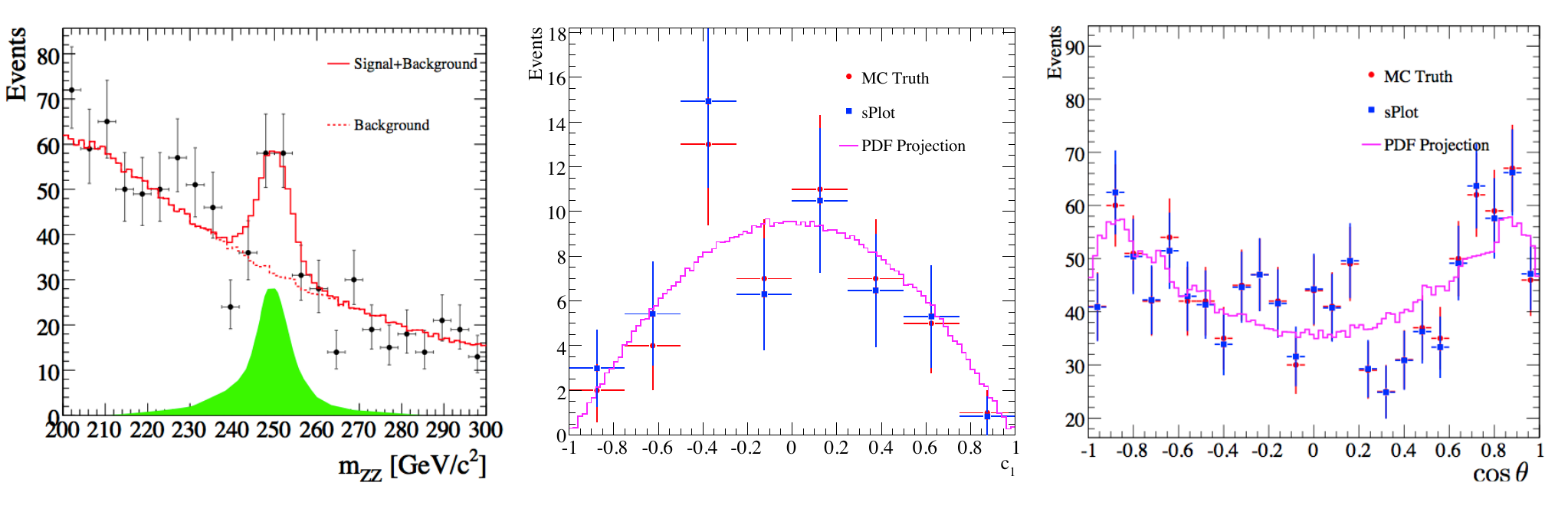}
\caption{Left: A signal on an $M(ZZ)$ distribution. Middle: sPlot of the $\cos\theta$ distribution
of the ``signal" events, compared with the Montecarlo truth and the (detector-shaped)
expected distribution, for $J^{PC}=0^{++}$. Right: Same as Middle, for the ``background"
events.
\label{sPlots}}
\end{center}
\end{figure}

There is a wonderful  ``s-Weighing" method for (much of) the exercise
of ascertaining the LHC's potential to select the preferred
hypothesis for an observed $H$ candidate.
Consider an $M[4\ell]$ distribution with an $H$ peak at
250 GeV, constructed with the standard expectations for signal and background, as in
Fig.~\ref{sPlots}. Performing a maximum-likelihood fit to this distribution
one can ascertain the probability of events in each $M[4\ell]$ bin to be signal
or background. Next one can astutely (and even statistically optimally) reweigh
the events into ``signal" and ``background" categories, to study their distributions
in other variables~\cite{ref:s-Plots}, such as $\cos\theta=\cos\theta_1$ or $\cos\theta_2$ 
in Fig.~\ref{sPlots}. In this pseudo-experiment one knows the ``Montecarlo truth",
compared in the figure with the impressive s-outcomes and the detector-shaped expectation.
We use the full (correlated) distributions in all mentioned variables, but $M_H$,
to confront ``data" with different hypothesis.

The astute reader has noticed that I have not mentioned the $\eta$ and $p_T$
distributions of the $ZZ$ or $ZZ^*$ pair (be it an $H$ signal or the irreducible background).
Event by event, one can undo the corresponding boost but,
to ascertain the detector-shaping effects, 
as in Fig.~\ref{Shaping},
for all the various SM or impostor $H$ objects,
one has to use a specific event generator. We have done it~\cite{ref:HLL}, 
but we chose to ``pessimize" our results in this respect, {\it not} exploiting the
 ($\eta,\, p_T$) distributions as part of the theoretical expectations
 (which for impostors would be quite model-dependent).
One reason is that the relevant parton distribution functions (PDFs)
will be better known by the time
a Higgs hunt becomes realistic. 
Another is that one can use the s-Weigh technique to extract and separately
plot the ($\eta,\,p_T$) distribution for signal and background. The production of a SM
$H$ -- but not that of most conceivable impostors --  is dominated by an extremely
theory-laden process: 
gluon fusion via a top loop. As a first step it is preferable 
*to see* whether or not the ($\eta,\,p_T$)
distribution of the s-sieved signal events
is that expected for $gg$ fusion, as opposed to $q\bar q$ annihilation\footnote{
The only impact of the difference between the two production processes is on the
detector-shaping effects. But these are not large enough for the
ensuing differences to affect our results.}.
The answer would be fascinating.

\section{Theory}
The most general Lorentz-invariant couplings of $J=0,1$ particles to
the polarization vectors $\epsilon_1^\mu$ and $\epsilon_2^\alpha$ of two $Z$s 
of  four-momenta $p_1$ and $p_2$ are given by the expressions:
  \begin{eqnarray}
  &&-i\,L_{\mu \alpha }= 
  X_0\, g_{\mu \alpha } 
  + (P_0+i \,Q_0) \,\epsilon _{\mu \alpha\sigma\tau }{p_1^\sigma p_2^\tau /M_Z^2}
  -(Y_0+i\, Z_0)\, {(p_1+p_2)_{\alpha } (p_1+p_2)_{\mu }/ M_Z^2}\, ,\nonumber\\
 &&-i\,L^{\rho \mu \alpha } =
 X_1 \left(g^{\rho
   \mu }\, p_1^{\alpha }\hspace*{-2pt}+\hspace*{-2pt}g^{\rho \alpha }\,
   p_2^{\mu }\right)+
       (P_1\hspace*{-2pt}+\hspace*{-2pt}i\, Q_1)\, \epsilon ^{\rho \mu \alpha\sigma}
       (p_1\hspace*{-2pt}-\hspace*{-2pt}p_2)_\sigma
       \nonumber
  \label{generalscalar}
  \end{eqnarray}
  The vertex for $J=2$ is cumbersome.
  The quantities $X_i,\, P_i...$  can be taken to be real, but for small absorptive effects. The
  expressions can be used to derive 
  the distribution functions  
  $pdf(J^{PC};\,M^*,\cos\Theta,\Phi,\cos\theta_1,\cos\theta_2,\varphi)$
   allowing one to determine the spin of an $H$ and the properties of the $HZZ$ coupling.  
 To give some $J=0$ examples: in the SM only $X_0=g\,M_Z/\cos\theta_W$ is nonvanishing.
 For $J=0^-$ only $Q_0\neq 0$. If $X_0$ and $Q_0$ (or $P_0$) $\neq 0$, the $HZZ$ vertex
 violates $P$ (or $CP$). For a ``composite scalar" $X_0,\,Y_0\neq 0$.

\section{Some results}

While Team 1 members are
 trying to establish the significance of the discovery of an object of specified
properties (as in Fig.~\ref{BRsDisco}, right), they may, with a few extra lines of code, be
extracting much more information from the same data set, by asking leading questions, 
NLQs, NNLQs..., whose answers are decreasingly statistically significant.

\begin{figure}[htb!]
\begin{center}
\includegraphics[width=1.0\textwidth]{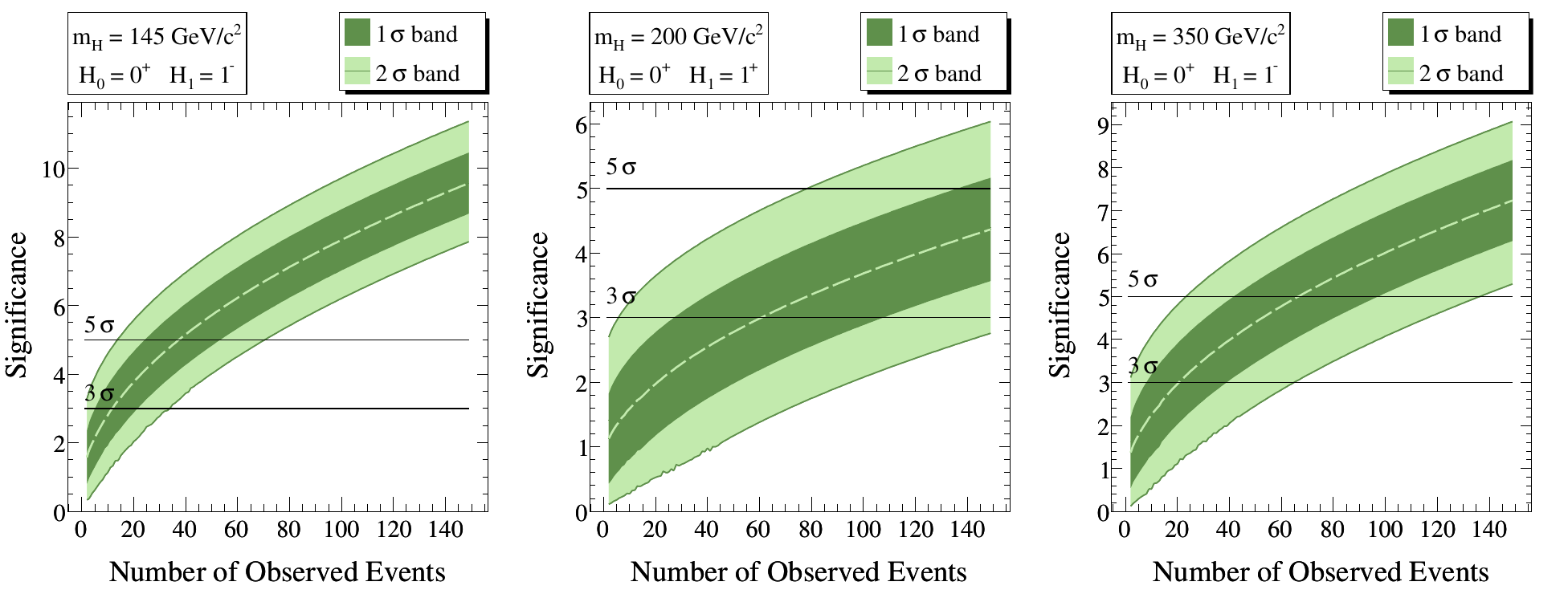}
\caption{Expected confidence levels, as functions of the number of events, 
to reject the wrong hypothesis ($H_0$,
the SM in this case)
in favour of the right one ($H_1$). Left and Right: $H_1$ is $1^-$, 
for $M_H=145$ and 350 GeV. Middle: $H_1$ is $1^+$, $M_H=200$ GeV.
\label{Results1}}
\end{center}
\end{figure}

\begin{figure}[htb!]
\begin{center}
\includegraphics[width=1.0\textwidth]{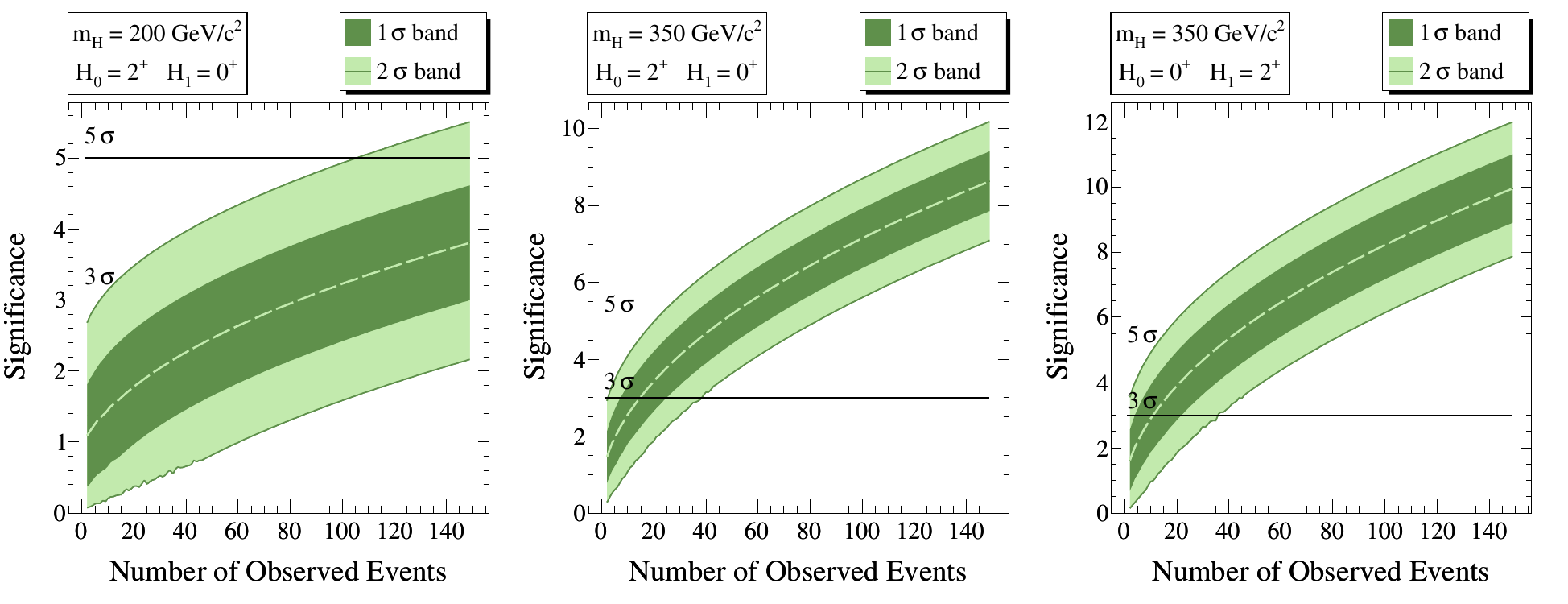}
\caption{Analogous to Fig.~\ref{Results1}, with the hypotheses $J^P=2^+$ and $0^+$,
once interchanged.
\label{Results2}}
\end{center}
\end{figure}

The quintessential LQ is which of two hypothesis describes the data best,
assuming that one of them is right. If the hypotheses are ``simple" (contain no parameters
to be fit) the Neyman-Pearson lemma guarantees
that the test is {\it universally most powerful}. 
Three examples are given in Fig.~\ref{Results1}. On its left and right
it is seen that it is ``easy"
(it takes a few tens
of events) to rule out the SM, if the observed resonance is an $M_H=145$ or 350 GeV 
vector. On its middle, we see that, if the object is an axial vector, it would be much harder.
This it is not due to the differing $J^P$, but to
the choice $M_H=200$ GeV. For masses close to the $H\to ZZ$ threshold, the level arm
provided by the lepton three-momenta is short, and the differences between {\it pdfs}
is diminished. In fact, as an answer to a NLQ, we have shown that,
except close to threshold, it is ``easy" to tell
any $J=0$ from any $J=1$ object, no matter how general their $HZZ$ couplings 
are~\cite{ref:HLL}. In Fig.~\ref{Results2} we see that it is easy,
if the SM is right, to exclude $J=2^+$ at $M_H=350$ GeV, but not at 200. We also see
that the interchange of right and wrong hypotheses leads to very similar expectations.

	\begin{figure}[htb!]
	\begin{center}
	\includegraphics[width=1.0\textwidth]{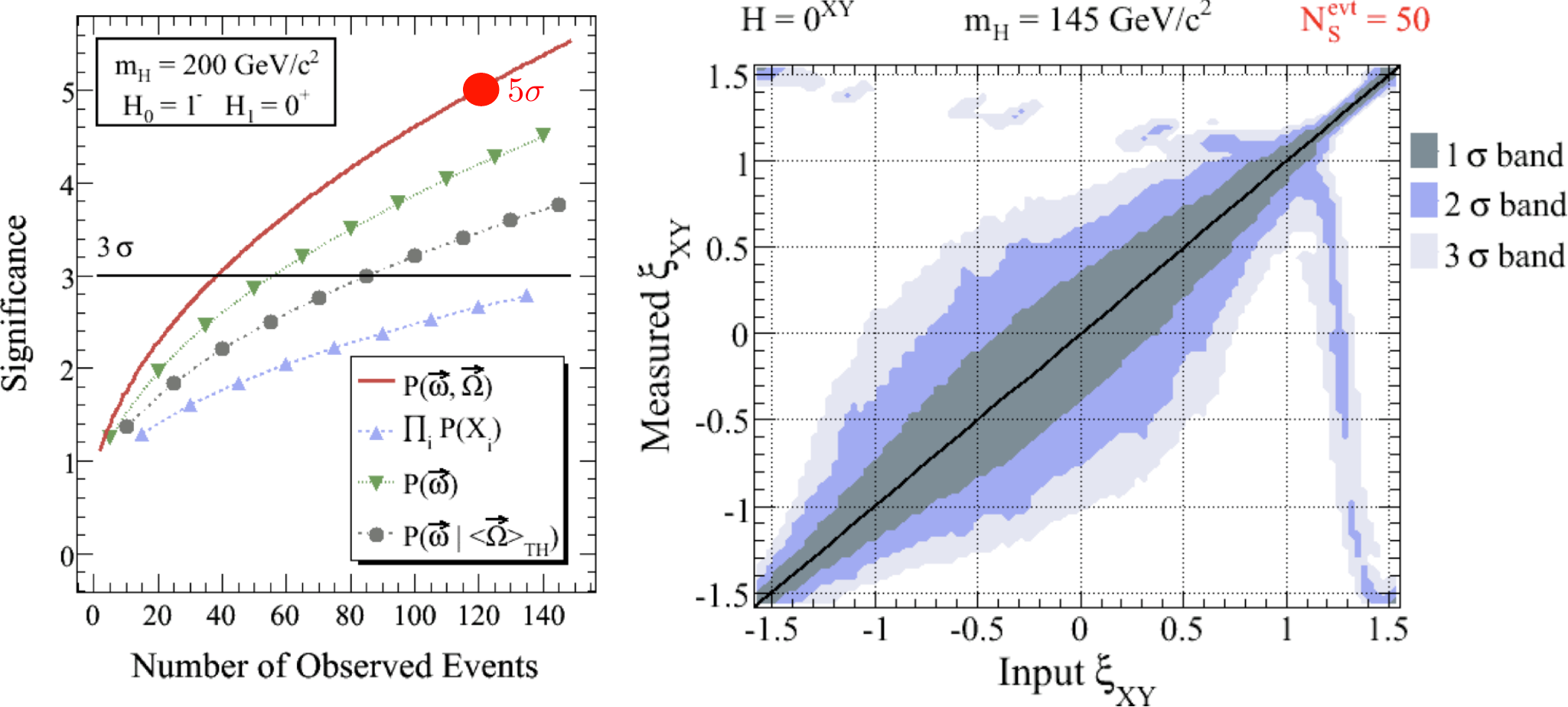}
	\caption{Left: Various choices of likelihood functions, 
	employing different sub-optimal sets of variables in their {\it pdfs,}
	 are compared with the choice containing all
	angular variables and their correlations (top-most curve). 
	Right: True and measured values
	of the mixing angle describing a composite scalar, for $M_H=145$ GeV.
	\label{Result}}
	\end{center}
	\end{figure}

On the right of Fig.~\ref{Result} is the answer to a NNLQ. We have assumed that
a composite $J^{PC}=0^{++}$ Higgs has been found and parametrized its $ZZ$ coupling
by an angle $\xi_{_{XY}}=\arctan(Y_0/X_0)$.  The measured value of 
$\xi_{_{XY}}$
is seen to be the input one, but for 50 events
the uncertainties on what the input was, to be read horizontally, are large. For this
case of a specific $J^{PC}$, but a complicated coupling, the various terms in the $pdf$
are not distinguishable on grounds of their properties under $P$ and $CP$. They do
strongly interfere for specific values of $\xi_{_{XY}}$,
and the results of  Fig.~\ref{Result} are not easy to obtain, requiring
a full Feldman-Cousins belt construction~\cite{ref:HLL}.

Given a small data set
constituting an initial discovery, one might settle for a stripped-down
analysis. The cost of such a sub-optimal choice is
shown on the left of Fig.~\ref{Result} for $M_H$$=$$200$ GeV,
illustrating the discrimination between the $0^{+}$ and
$1^{-}$ hypotheses for likelihood definitions that exploit different
sets of variables. 
N-dimensional {\it pdfs} in the variables
$\{a_{1},\cdots,a_{N}\}$ are denoted $P(a_{1}, \cdots ,a_{N})$, while
$\prod_{i} P(X_{i})$ is constructed
from one-dimensional {\it pdfs} for all variables, ignoring
(erroneously) their correlations. $P(\vec{\omega}\, |
\langle\vec{\Omega}\rangle_{\rm TH})$ are {\it pdfs} including the
variables $\vec{\omega}$ and their correlations, but with the
hypothesis $1^{-}$ represented by a {\it pdf} in which the variables
$\vec{\Omega}$ have been integrated out.
  The likelihood $P(\vec{\omega}\, | \langle\vec{\Omega}\rangle_{\rm
  TH})$ performs badly even relative to $P(\vec{\omega})$, which uses
fewer angular variables. The two differ only in that the first
construction implicitly assumes a uniform $4\pi$ coverage of the
observed leptons (an assumption customary in the literature) as if the
muon $p_{T}$ and $\eta$ analysis requirements did not depend on the
$\vec\Omega$ angular variables. 

\begin{figure}[htb!]
\begin{center}
\includegraphics[width=1.0\textwidth]{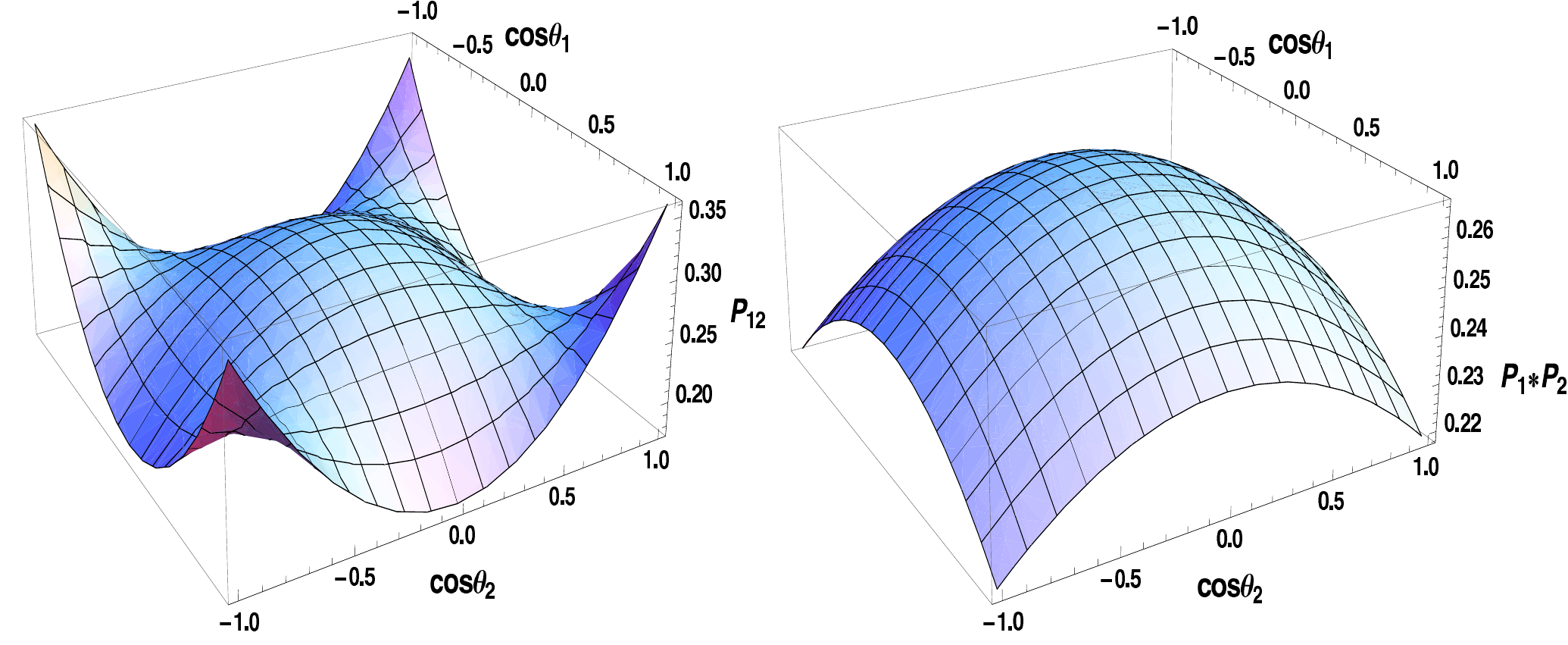}
\caption{The {\it pdfs} of the SM at $M_H=200$ GeV, integrated in all variables
but $\cos\theta_1$ and $\cos\theta_2$. Left: the correct  
$P(\cos\theta_1,\cos\theta_2)$. Right: the  ``approximation" 
$P(\cos\theta_1)\times P(\cos\theta_2)$.
\label{Results3}}
\end{center}
\end{figure}

Treating the correlated angular variables as
uncorrelated, as in the $\prod_{i} P(X_{i})$ example of Fig.~\ref{Result}, 
not only degrades the discrimination significance but
would lead to time-dependent, ultimately wrong conclusions.
Assume, for example, the SM with $m_H = 200$
GeV.  Let the data be fit to either a fully
correlated {\it pdf} or an uncorrelated one. The projections
of the corresponding theoretical {\it pdfs}, involving only the variables 
cos$\,\theta_1$ and cos$\,\theta_2$, are illustrated in Fig.~\ref{Results3}. 
On the left (right) of the figure we see $P[{\rm cos}\,\theta_1,\,{\rm cos}\,\theta_2]$
($P[{\rm cos}\,\theta_1]$$\times$$P[{\rm cos}\,\theta_2]$). With limited statics -- insufficient
to distinguish between the correlated and uncorrelated distributions --
the correct conclusion will be reached: the data are compatible with the SM.
But, as the statistics are increased, the data will significantly deviate from
the $P[{\rm cos}\,\theta_1]\times P[{\rm cos}\,\theta_2]$ distribution, and a false
rejection of the SM hypothesis would become increasingly supported.

The difference between $P[{\rm cos}\,\theta_1,\,{\rm cos}\,\theta_2]$
and $P[{\rm cos}\,\theta_1]$$\times$$P[{\rm cos}\,\theta_2]$ is {\it precisely}
what an unbelieving Einstein called {\it spooky action at a distance}. But,
mercifully for physicists, the Lord
is subtle *and* perverse.

\section{Conclusions}

I have alleged, by way of example, that for a fixed detector performance and
integrated luminosity (and no extra Swiss Francs) it pays to have {\it ab initio}
an analysis combining discovery and scrutiny. This is arguably true for many
physics items other than $H\to 4\ell$. They readily come to mind.

\end{document}